\documentclass{aa}
\usepackage{graphicx}
\usepackage{amsmath}
\usepackage{txfonts}
\usepackage{subfigure}

\usepackage{natbib}
\bibpunct{(}{)}{;}{a}{}{,} 

\usepackage{hyperref}

\graphicspath{{figures/}}

\title{Shear mixing in stellar radiative zones}
\subtitle{II.~Robustness of numerical simulations}
\titlerunning{Shear mixing in stellar radiative zones. II.~Robustness of numerical simulations}
\author{V. Prat\inst{1,2} \and J. Guilet\inst{1} \and M. Viallet\inst{1} \and E. M\"uller\inst{1}}
\institute{Max-Planck Institut f\"ur Astrophysik, Karl-Schwarzschild-Str.~1, 85748, Garching bei M\"unchen, Germany
\and Laboratoire AIM Paris-Saclay, CEA/DRF - CNRS - Universit\'e Paris Diderot, IRFU/SAp Centre de Saclay, F-91191 Gif-sur-Yvette, France}
\date{Received ? / Accepted ?}

\begin{document}

\abstract
{Recent numerical simulations suggest that the model by Zahn (1992, \aap, 265, 115) for the turbulent mixing of chemical elements due to differential rotation in stellar radiative zones is valid.}
{We investigate the robustness of this result with respect to the numerical configuration and Reynolds number of the flow.}
{We compare results from simulations performed with two different numerical codes, including one that uses the shearing-box formalism.
We also extensively study the dependence of the turbulent diffusion coefficient on the turbulent Reynolds number.}
{The two numerical codes used in this study give consistent results.
The turbulent diffusion coefficient is independent of the size of the numerical domain if at least three large turbulent structures fit in the box.
Generally, the turbulent diffusion coefficient depends on the turbulent Reynolds number.
However, our simulations suggest that an asymptotic regime is obtained when the turbulent Reynolds number is larger than $10^3$.}
{Shear mixing in the regime of small P\'eclet numbers can be investigated numerically both with shearing-box simulations and simulations using explicit forcing.
Our results suggest that Zahn's model is valid at large turbulent Reynolds numbers.}


\maketitle
\section{Introduction}

Stellar evolution plays a central role in astrophysics by linking observable quantities to fundamental parameters of stars, such as mass, radius, and age.
These parameters also provide constraints on galaxies, which are made of stars, and on planetary systems, which orbit around them.
In addition, stellar models are used to calibrate many measurements for other domains of astrophysics, such as cosmological distance measurements based on models of Cepheids and type-Ia supernovae.

Stellar evolution theory involves many physical processes with very different lengths and occuring at very different time scales.
In particular, rotationally induced transport processes, as well as overshoot and various magneto-hydrodynamical (MHD) instabilities, often refered to as ``extra mixing'', have a significant impact on the distribution of angular momentum and chemical elements, and thus on the structure and the evolution of stars.
The predictive power of stellar evolution codes is limited by large uncertainties related to our poor knowledge of these (magneto-)hydrodynamical processes and by the fact that, being fundamentally 3D, they cannot be included self-consistently in 1D computations.
Some evolution codes include rotation \citep{Meynet} or other 3D processes, such as overshoot \citep{Moravveji}, semi-convection \citep{Ding}, thermohaline mixing \citep{Wachlin}, or the magneto-rotational instability \citep{Wheeler}.
However, these processes, including convection itself, are described by phenomenological models which involve free parameters.

Another problem of stellar evolution is the fact that all the information we have about stars comes from their surfaces (except for the Sun where we also have information from neutrinos) and is generally not spatially resolved.
This implies that there is no straightforward way to constrain the transport generated by hydrodynamical processes inside stars.
However, we do have some constraints, mainly from spectroscopy and asteroseismology.

First, the analysis of absorption lines in stellar spectra gives access to surface chemical abundances of stars.
Some of these abundances are strongly affected by mixing processes occurring in certain regions of stars.
For instance, elements produced (helium and nitrogen) or depleted (carbon) in the core during the CNO cycle are good indicators of mixing between the core and the surface \citep[see for example][]{Hunter, Martins}.
One major issue is that measured abundances depend on the initial chemical composition of observed stars, which is usually not known.

In contrast, asteroseismology allows us to probe the interiors of stars by analysing their oscillation spectra.
Indeed, regular spectral patterns can be related to the stellar structure, including the rotation profile.
These techniques have been used so far mainly for the Sun and a few red giants and subgiants \citep{Beck, Mosser, Deheuvels2012,Deheuvels2014, Deheuvels2015}, but a great quantity of new data from other stars is now available thanks to the CoRoT and Kepler satellites.
This is clearly a very promising source of additional constraints on transport processes \citep[see for example][]{Moravveji}.

The so-called shear mixing is one of these processes.
It occurs when differential rotation is strong enough to overcome the effect of stable stratification in a radiative zone.
Stellar radiative zones are characterised by a very efficient radiative thermal diffusion, which tends to weaken the stabilising effect of thermal stratification.
This has been formalised by \citet{Zahn} using phenomenological arguments.
Since then, several related models including additional physics such as chemical stratification or horizontal turbulence have been proposed and implemented in stellar evolution codes \citep{MaederMeynet, TalonZahn}.
More rigourous models based on mean-field equations exist.
However, these models usually express the mixing rate as a function of quantities which are not available in stellar evolution codes \citep[e.g.][]{LindborgBrethouwer}, and additional assumptions such as closure models \citep[e.g.][]{Canuto1998, Canuto2011} are thus required, introducing uncertainties.
Although it would be interesting to test such more complex models with numerical simulations, we focus here on Zahn's model, which is commonly used in stellar evolution calculations.

An increasing number of studies suggest that the transport predicted by Zahn's model (or related ones) is insufficient to fit some obervational constraints.
For example, synthetic stellar populations computed by \citet{Brott} and \citet{Potter} cannot reproduce several groups of stars observed by \citet{Hunter}, notably nitrogen-rich slow rotators.
Note that the existence of these stars might be explained by rotational mixing if some braking is active at the surface, for instance due to a surface magnetic field \citep{Meynet2011}.
Other studies show that additional mixing is needed to explain observed rotation profiles of red giants or subgiants \citep{Eggenberger, Ceillier, Marques}.
Moreover, even adding other transport processes is still not sufficient to explain the observations (\citealt{Cantiello}, for the Tayler-Spruit dynamo; \citealt{Fuller, Belkacem}, for internal gravity waves).
Constraints from the spin of neutron stars and white dwarfs at birth also indicate that the transport as predicted by Zahn's model is not sufficient (\citealt{Heger}, for neutron stars; \citealt{Suijs}, for white dwarfs).
Additional constraints are needed to determine whether these discrepancies point to an incorrect description of shear mixing or to the presence of another transport process.  

Because of the degeneracy of observational constraints and the impossibility of performing laboratory experiments in the stellar regime, numerical simulations are a second way to obtain new constraints on specific transport processes.
The purpose of this work is to test Zahn's family of models for shear mixing with local direct numerical simulations, and potentially to propose new prescriptions including new physical ingredients.
Whereas in the previous papers \citep[][thereafter refered to as PL13 and Paper I, respectively]{PratLignieres2013, PratLignieres2014} we focused on thermal diffusion and the dynamical effect of chemical stratification, the present paper is intended to test the robustness of the constraints on Zahn's model that one can obtain from shear mixing simulations.
We achieve this by comparing the results of simulations performed with two different numerical codes and different radial boundary conditions, and by studying the effect of viscosity and the size of the numerical domain.

The paper is organised as follows.
We summarise the model of shear mixing proposed by \citet{Zahn} in Sect.~\ref{sec:zahn}, and we describe the two numerical codes used in this paper in Sect.~\ref{sec:configs}.
The results of our simulations are presented in Sect.~\ref{sec:results}.
Finally, we discuss the robustness of numerical constraints on Zahn's model in Sect.~\ref{sec:discuss}.

\section{Model by \citet{Zahn}}
\label{sec:zahn}

In the laminar, adiabatic, non-viscous case, the classical Richardson criterion for linear shear instability reads \citep{Miles}
\begin{equation}
Ri < \frac14,
\end{equation}
where $Ri=N^2/S^2$ is the Richardson number comparing the stabilising effect of thermal stratification (given by the Brunt-V\"ais\"al\"a frequency $N$) and the destabilising effect of differential rotation (given by the shear rate $S={\rm d}U/{\rm d}z$ of a velocity field $U(z)$ varying along the spatial coordinate $z$).
Typical Richardson number values in stellar radiative zones are between $10^2$ and $10^4$.
Following \citet{Townsend}, \citet{Zahn1974} introduced the effect of thermal diffusion on thermal stratification.
The main assumptions underlying Zahn's model are:
(i) in the regime of very high thermal diffusivities, the Richardson criterion is replaced by
\begin{equation}
RiPe_\ell < R_{\rm crit},
\end{equation}
where $Pe_\ell=w\ell/\kappa$ is the turbulent P\'eclet number (which can be seen as the ratio between the diffusive cooling time and the dynamical time of turbulence), $w$ and $\ell$ turbulent velocity and length scales, $\kappa$ the thermal diffusivity of the fluid, and $R_{\rm crit}$ the critical value of $RiPe_\ell$, of the order of one;
(ii) turbulent flows tend to reach a statistical steady state which is marginally stable (i.e. $RiPe_\ell=R_{\rm crit}$);
(iii) the turbulent diffusion coefficient $D_{\rm t}$ is proportional to $w\ell$; and
(iv) the turbulent Reynolds number (i.e. the ratio between the viscous time and the dynamical time of turbulence, which typically is between $10^2$ and $10^5$ in stellar radiative zones) $Re_\ell=w\ell/\nu$, where $\nu$ is the viscosity of the fluid, is larger than a critical value $Re_{\rm crit}$ of the order of $10^3$.
These assumptions, especially the first one, which arises from the hypothesis that a linear treatment can provide some approximate treatment of turbulent flows, are not rigourously justified, and hence need to be tested with numerical simulations such as those presented in this work.

From the first two assumptions, we can deduce
\begin{equation}
Pe_\ell=\frac{R_{\rm crit}}{Ri},
\end{equation}
and by definition of the turbulent P\'eclet number, we have
\begin{equation}
    \label{eq:ul}
    w\ell = \frac{\kappa R_{\rm crit}}{Ri}.
\end{equation}
Finally, the third assumption gives us
\begin{equation}
    \label{eq:dtzahn}
    D_{\rm t}\propto \frac{\kappa}{Ri}.
\end{equation}

It is important to note that in the asymptotic regime described by \citet{Zahn}, the turbulent diffusion coefficient does not depend on viscosity.
However, the existence of the fourth assumption suggests that when it is not verified, the turbulent diffusion coefficient may depend on the value of the turbulent Reynolds number.
In the asymptotic regime, from Eq.~\eqref{eq:ul}, we get
\begin{equation}
    \label{eq:rel_ripr}
    Re_\ell \propto (RiPr)^{-1},
\end{equation}
where $Pr=\nu/\kappa$ is the Prandtl number (comparing the relative effects of viscosity and thermal diffusion), which is typically of the order of $10^{-6}$ or smaller in stellar interiors.
This suggests that $RiPr$ is a proxy for the turbulent Reynolds number.
We can thus describe the dependence of the turbulent diffusion coefficient on the turbulent Reynolds number as a dependence on $RiPr$.
In particular, one can check how the quantity $D_{\rm t}/(\kappa Ri^{-1})$, which is constant in Zahn's model according to Eq.~\eqref{eq:dtzahn}, varies with $RiPr$.
This will be done in Sect.~\ref{sec:rel}.
Such a dependence is easier to implement in stellar evolution codes, where $RiPr$ can be computed, but not $Re_\ell$.

\section{Physical and numerical configurations}
\label{sec:configs}

The simulations presented in this paper have been performed with two different codes.
The first one is the Bala\"itous code, that we used in PL13 and Paper~I.
This corresponds to what we call the forcing configuration.
The second one is the Snoopy code \citep{LesurLongaretti}, associated with the shearing-box configuration.

The physical configurations used in this paper are essentially the same, as in PL13 and Paper I, and the two codes also have some numerical aspects in common, which are listed in Sect.~\ref{sec:common}.
The forcing configuration is described in Sect.~\ref{sec:forcing} and we present the specificities of the shearing-box configuration in Sect.~\ref{sec:shear}.

\subsection{Common aspects}
\label{sec:common}

\begin{figure}
    \resizebox{\hsize}{!}{\includegraphics{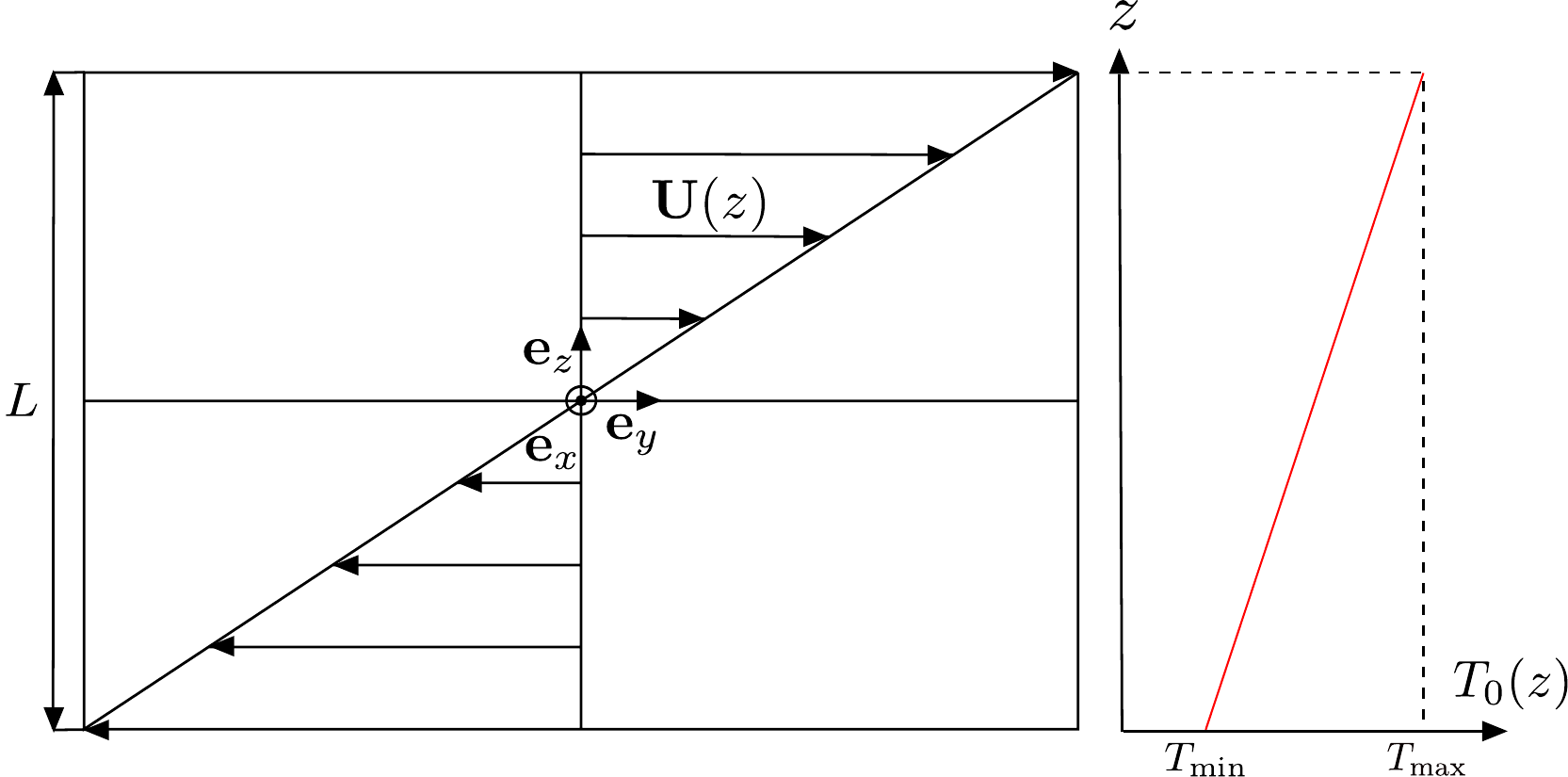}}
    \caption{Sketch of the flow configuration}
    \label{fig:flow}
\end{figure}

All our simulations are performed using the Boussinesq approximation, which neglects density fluctuations except in the buoyancy term, and the so-called small-P\'eclet-number approximation \citep[SPNA, see][]{Lignieres}, which allows us to explore the regime of very high thermal diffusivities at a reasonable computational cost.
Local mean gradients are characterised by a uniform velocity shear, and uniform temperature and chemical composition gradients in the vertical/radial direction, as described in Fig.~\ref{fig:flow}.

We choose $L$, $S^{-1}$, $\Delta U=SL$, $\Delta T=L{\rm d}T/{\rm d}{z}$, $\Delta c=L{\rm d}C/{\rm d}{z}$, and $\rho S^2L^2$ as length, time, velocity, temperature, concentration, and pressure units (where $L$ is the size of the numerical domain, $T(z)$ and $C(z)$ are the background temperature and concentration profiles, and $\rho$ is the background density).
In this framework, the dimensionless Navier-Stokes equations read
\begin{align}
\vec\nabla\cdot\vec u                                           &= 0,\\
\frac{\partial\vec u}{\partial t}+(\vec u\cdot\vec\nabla)\vec u &= -\vec\nabla P+RiPe\,\psi\vec e_z+\frac{1}{Re}\Delta\vec u, \label{eq:momentum}\\
u_z                                                             &= \Delta\psi,
\end{align}
with $\psi=\theta/Pe$, where $\vec u$, $P$, and $\theta$ are the dimensionless velocity, pressure, and temperature fluctuations around the background profile, and $Pe=SL^2/\kappa$ is the P\'eclet number, assumed to be very small. 
These equations highlight the two main control parameters: (i) the product of the Richardson and P\'eclet numbers $RiPe=N^2L^2/(S\kappa)$ (where $N^2=\alpha g\Delta T/L$, $\alpha$ being the thermal expansion coefficient of the fluid and $g$ the local gravity), also called the Richardson-P\'eclet number, which characterises the effect of stratification affected by thermal diffusion, and (ii) the Reynolds number $Re=SL^2/\nu$, which characterises the effect of viscosity.
These control parameters are clearly dependent on the size of the numerical domain $L$.
One can define their turbulent counterparts $RiPe_\ell=N^2w\ell/(S^2\kappa)$ and $Re_\ell=w\ell/\kappa$, based on the turbulent velocity and length scales, which are diagnostic variables.
The only relevant control parameter that does not depend on the size of the numerical domain is the product of the Richardson and Prandtl numbers $RiPr=RiPe/Re=N^2\nu/(S^2\kappa)$.

To these equations, we add the equation for passive advection/diffusion of the chemical concentration:
\begin{equation}
    \label{eq:advdiff}
    \frac{\partial c}{\partial t}+\vec u\cdot\vec\nabla c=\frac{1}{Pe_{\rm c}}\Delta c,
\end{equation}
which introduces a new control parameter, the chemical P\'eclet number $Pe_{\rm c}=SL^2/D_{\rm m}$, which characterises the effect of the chemical molecular diffusity $D_{\rm m}$.

Numerically, both codes use a Fourier colocation method associated with periodic boundary conditions in the horizontal ($x$- and $y$-) directions.
They also both allow us to perform direct numerical simulations (DNS), in which all the physically relevant scales of the problem are resolved, typically down to the dissipation scale.
The main differences between the codes lie in the numerical method and the boundary conditions used in the vertical ($z$-) direction, and in the way background profiles (of shear, temperature, and composition) are imposed.

\subsection{Forcing configuration}
\label{sec:forcing}

The forcing configuration uses compact finite differences in the vertical direction, associated with impenetrable boundary conditions and imposed shear, temperature and chemical composition at the upper and lower boundaries.
Due to these boundary conditions, the flow is not rigourously statistically homogeneous in the vertical direction close to the upper and lower boundaries, as mentioned in PL13 and Paper I.

The linear background profiles are imposed by forcing terms $\vec f_{\rm v}$ and $f_{\rm c}$ added to Eqs.~\eqref{eq:momentum} and~\eqref{eq:advdiff}, respectively.
These terms are defined as
\begin{equation}
    \vec f_{\rm v} = \frac{\vec U(z)-\overline{\vec u}}{\tau}\quad   \text{and}  \quad f_{\rm c} = \frac{C(z)-\overline{c}}{\tau},
\end{equation}
where $\vec U(z)$ and $C(z)$ are the background velocity and composition profiles, the overline denoting the horizontal average, and $\tau$ is the forcing time scale.
Normally a similar term should be added to the temperature equation, but in the SPNA, only infinitesimally small temperature fluctuations are considered.

Tests have shown that, provided that the forcing time scale $\tau$ is much smaller than a few hundred shear times, its exact value does not affect the results. 
As a consequence of the forcing method, the horizontally-averaged profiles (here $\overline{\vec u}$ and $\overline{c}$) remain very close to the background profiles (here $\vec U$ and $C$).

\subsection{Shearing-box configuration}
\label{sec:shear}

The shearing-box configuration uses a Fourier colocation method also in the vertical direction, associated with shear-periodic boundary conditions at the upper and lower boundaries.
This is made possible by the fact that the equations are solved for the fluctuations of all quantities around their background profiles and not for the quantities themselves.
The resulting flow is rigourously statistically homogeneous in all directions, in constrast with the forcing configuration.

There is no explicit forcing in the shearing-box configuration, but the fact that fluctuations are shear-periodic in the vertical direction ensures that the spatial averages of the shear and of the temperature and composition gradients are equal to their background values.
Equivalently, it means that the horizontal averages of velocity, temperature and composition differences between the upper and lower boundaries are fixed.
Consequently, horizontally averaged gradients are not necessarily uniform at a given time and mean profiles are able to evolve under the action of the instability.

\section{Results of the simulations}
\label{sec:results}

We performed five series of simulations at constant chemical P\'eclet number $Pe_{\rm c}=4\cdot10^4$, with different values of $RiPe$: one with Bala\"itous at $Re=4\cdot10^4$, and four with Snoopy with $Re$ ranging from $2\cdot10^4$ to $1.6\cdot10^5$.
In each simulation, we estimate the turbulent diffusion coefficient as
\begin{equation}
D_{\rm t} = -\frac{\langle c'u_z\rangle}{{\rm d}C/{\rm d}z},
\end{equation}
where $\langle\rangle$ denotes the temporal and spatial average, assumed to be equivalent to the statistical average, $c'$ is the fluctuation of chemical composition about its background profile, and $u_z$ is the vertical component of the velocity.
One can also estimate the turbulent viscosity coefficient, also known as eddy viscosity, as
\begin{equation}
    \nu_{\rm t} = -\frac{\langle u'_yu_z\rangle}{{\rm d}U/{\rm d}z},
\end{equation}
where $u'_y$ is the fluctuation of the $y$-component of the velocity about the mean shear profile.
The SPNA simulation described in PL13 is also considered here for comparison.
Table~\ref{tab:res} summarises the relevant quantities of our simulations.
\begin{table*}
    \caption{Summary of our simulations}
    \label{tab:res}
    \centering
    \(
    \begin{array}{ccccccccccc}
    \hline\hline
    \#                      &   Re              &   RiPe    &   RiPr            &   D_{\rm t}/(\kappa Ri^{-1})  &   \ell/L  &   Re_\ell         &   RiPe_\ell   &   D_{\rm t}/(w\ell)   &   D_{\rm t}/\nu   &   \nu_{\rm t}/\nu \\
                            &   (\times10^4)    &           &   (\times10^{-4}) &   (\times10^{-2})             &           &   (\times10^2)    &               &   (\times10^{-2})     &                   &                   \\\hline
    \multicolumn{11}{c}{\text{Bala\"itous}}    \\
    1\tablefootmark{d}      &   4               &   20      &   5               &   4.07                        &   0.118   &   5.06            &   0.253       &   16.1                &   81.5            &   66.0            \\
    2\tablefootmark{ad}     &   4               &   50.8    &   12.7            &   5.58                        &   0.111   &   3.35            &   0.426       &   13.1                &   42.7            &   33.6            \\
    3\tablefootmark{d}      &   4               &   140     &   35              &   6.43                        &   0.107   &   2.22            &   0.776       &   8.29                &   18.4            &   16.4            \\
    4\tablefootmark{d}      &   4               &   240     &   60              &   4.52                        &   0.116   &   1.68            &   1.01        &   4.49                &   7.54            &   7.54            \\\hline
    \multicolumn{11}{c}{\text{Snoopy}}   \\
    5\tablefootmark{e}      &   2               &   4       &   2               &   2.45                        &   0.474   &   19.1            &   0.381       &   6.42                &   122             &   102             \\
    6\tablefootmark{e}      &   2               &   8       &   4               &   3.36                        &   0.422   &   13.5            &   0.540       &   6.22                &   84.0            &   65.9            \\
    7                       &   2               &   20      &   10              &   4.90                        &   0.322   &   7.43            &   0.743       &   6.60                &   49.0            &   37.6            \\
    8                       &   2               &   40      &   20              &   6.09                        &   0.277   &   5.15            &   1.03        &   5.92                &   30.5            &   24.8            \\
    9                       &   2               &   80      &   40              &   6.03                        &   0.243   &   3.36            &   1.34        &   4.49                &   15.1            &   13.7            \\
    10                      &   2               &   120     &   60              &   4.03                        &   0.233   &   2.34            &   1.41        &   2.87                &   6.71            &   6.78            \\
    11\tablefootmark{b}     &   2               &   140     &   70              &   -                           &   -       &   -               &   -           &   -                   &   -               &   -               \\
    12\tablefootmark{de}    &   4               &   4       &   1               &   2.33                        &   0.484   &   39.7            &   0.397       &   5.88                &   233             &   202             \\
    13\tablefootmark{de}    &   4               &   8       &   2               &   2.68                        &   0.406   &   24.0            &   0.480       &   5.59                &   134             &   107             \\
    14\tablefootmark{d}     &   4               &   20      &   5               &   4.01                        &   0.296   &   12.6            &   0.639       &   6.38                &   80.3            &   60.9            \\
    15\tablefootmark{d}     &   4               &   40      &   10              &   5.07                        &   0.231   &   7.75            &   0.775       &   6.55                &   50.7            &   39.5            \\
    16\tablefootmark{d}     &   4               &   60      &   15              &   5.77                        &   0.208   &   6.14            &   0.920       &   6.27                &   38.4            &   30.9            \\
    17\tablefootmark{d}     &   4               &   80      &   20              &   6.16                        &   0.193   &   5.14            &   1.03        &   5.98                &   30.8            &   25.5            \\
    18\tablefootmark{d}     &   4               &   200     &   50              &   5.57                        &   0.167   &   2.91            &   1.46        &   3.83                &   11.1            &   10.6            \\
    19\tablefootmark{d}     &   4               &   240     &   60              &   4.35                        &   0.164   &   2.43            &   1.46        &   2.99                &   7.26            &   7.29            \\
    20\tablefootmark{b}     &   4               &   280     &   70              &   -                           &   -       &   -               &   -           &   -                   &   -               &   -               \\
    21\tablefootmark{e}     &   8               &   8       &   1               &   2.81                        &   0.414   &   50.9            &   0.509       &   5.52                &   281             &   227             \\
    22                      &   8               &   20      &   2.5             &   3.33                        &   0.290   &   23.2            &   0.580       &   5.74                &   133             &   102             \\
    23                      &   8               &   40      &   5               &   4.02                        &   0.210   &   12.8            &   0.639       &   6.30                &   80.5            &   62.5            \\
    24                      &   8               &   80      &   10              &   5.03                        &   0.163   &   7.85            &   0.785       &   6.41                &   50.3            &   40.4            \\
    25                      &   8               &   200     &   25              &   6.23                        &   0.132   &   4.61            &   1.15        &   5.41                &   24.9            &   21.7            \\
    26                      &   8               &   400     &   50              &   5.46                        &   0.126   &   3.12            &   1.56        &   3.49                &   10.9            &   10.6            \\
    27                      &   8               &   500     &   62.5            &   4.17                        &   0.143   &   2.96            &   1.85        &   2.25                &   6.67            &   6.80            \\
    28\tablefootmark{b}     &   8               &   600     &   75              &   -                           &   -       &   -               &   -           &   -                   &   -               &   -               \\
    29\tablefootmark{c}     &   16              &   16      &   1               &   3.67                        &   0.358   &   70.8            &   0.708       &   5.18                &   367             &   296             \\
    30\tablefootmark{c}     &   16              &   40      &   2.5             &   3.59                        &   0.223   &   26.9            &   0.672       &   5.34                &   144             &   114             \\\hline
    \end{array}
    \)
    \tablefoot{
        For numerical reasons the definition of $\ell$ is not the same in the two codes.
        This may explain the observed discrepancies for $\ell/L$, $Re_\ell$, $RiPe_\ell$, and $D_{\rm t}/(w\ell)$ between the codes.\\
        \tablefoottext{a}{SPNA simulation from PL13}
        \tablefoottext{b}{Simulations without turbulence}
        \tablefoottext{c}{Snoopy simulations with $512^3$ grid points instead of $256^3$}
        \tablefoottext{d}{Simulations considered in Sect.~\ref{sec:comp}}
        \tablefoottext{e}{Snoopy simulations considered as not relevant according to the criterion on $\ell/L$ given in Sect.~\ref{sec:size}}
    }
\end{table*}

In this section, we investigate four different aspects of our simulations.
First, we compare the results given by the two codes in Sect.~\ref{sec:comp}.
Second, we analyse the effect of the limited size of the numerical domain in Sect.~\ref{sec:size}.
Third, we study the dependence of the turbulent transport on the turbulent Reynolds number in Sect.~\ref{sec:rel}.
Finally, we constrain the relation between the coefficients of turbulent diffusion and viscosity in Sect.~\ref{sec:nul}.

\subsection{Comparison of the two codes}
\label{sec:comp}

\begin{figure}
    \resizebox{\hsize}{!}{\includegraphics{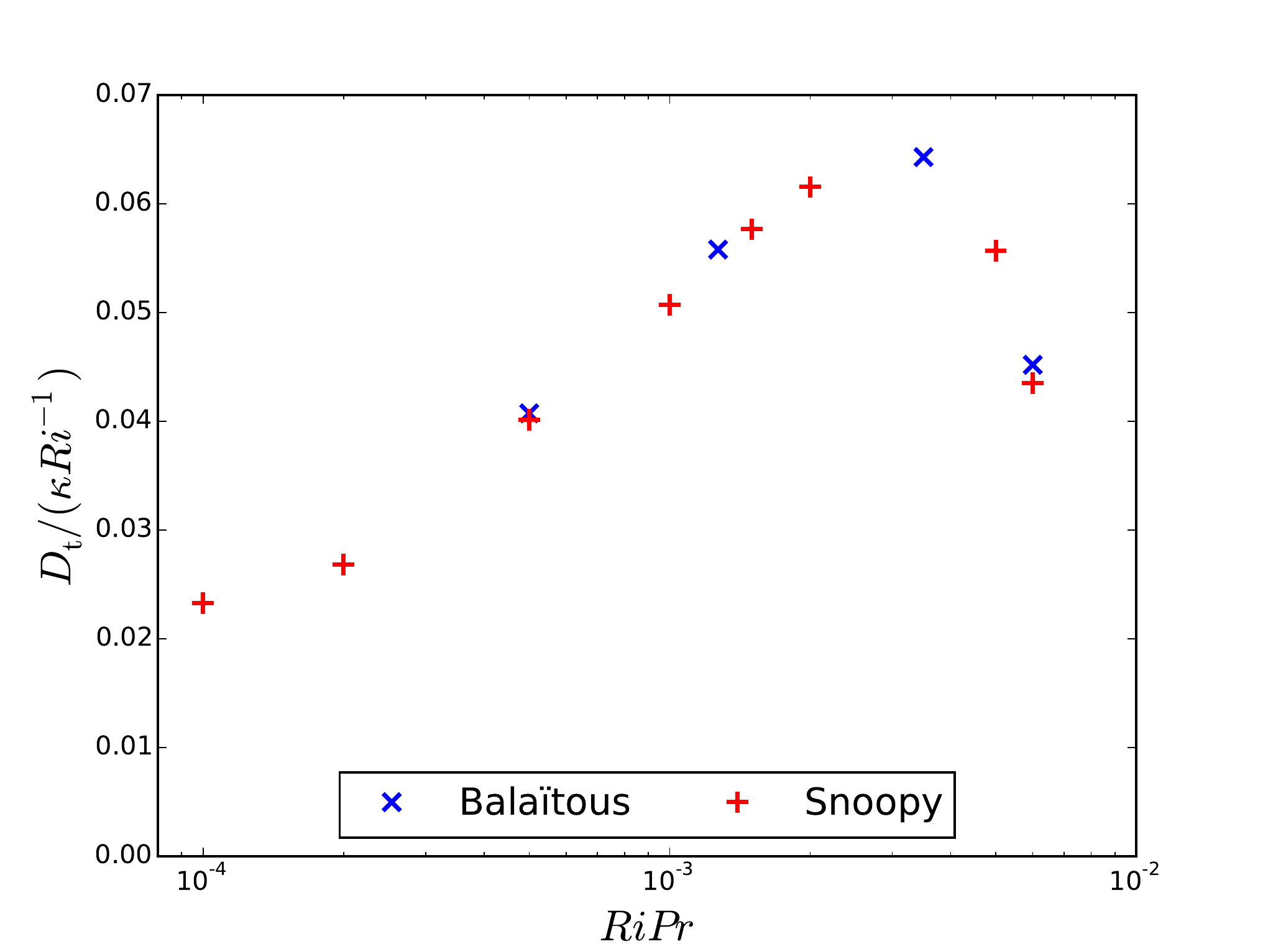}}
    \caption{$D_{\rm t}/(\kappa Ri^{-1})$ as a function of $RiPr$ for simulations with $Re=4\cdot10^4$.}
    \label{fig:comp}
\end{figure}

To compare the behaviour of the two codes in a similar configuration, we consider here only simulations at $Re=4\cdot10^4$ (indicated in Table~\ref{tab:res} by a superscript \emph{d}).
Figure~\ref{fig:comp} shows the quantity $D_{\rm t}/(\kappa Ri^{-1})$ as a function of $RiPr$ for the two codes.
One can clearly see that the two codes give very similar results.
For values of $RiPr$ where both codes have been tested, the relative difference between the two estimated transport coefficients never exceeds a few percent.
For values of $RiPr$ where only one code has been tested, our results generally follow the same curve.
This indicates that both codes are consistent for the study of turbulent transport in the regime of small P\'eclet numbers.
We thus validate the shearing-box formalism for the study of the shear instability in this regime.
This is an encouraging result, since the Snoopy code is more efficient and scalable than Bala\"itous, mainly because of the 3D fast Fourier transform.

One may notice in Table~\ref{tab:res} that for the simulations considered in this section, the values of $\ell/L$, $Re_\ell$, $RiPe_\ell$, and $D_{\rm t}/(w\ell)$ significantly differ between the codes.
This results at least partly from the fact that the length $\ell$ is differently defined in the two codes.
In Bala\"itous, $\ell$ is the integral scale of turbulence based on horizontal 2D spectra, whereas in Snoopy it is the integral scale based on 3D spectra.
As shear flows are expected to be anisotropic, it is not surprising that these two definitions give different results.
To summarise, we can write
\begin{equation}
\ell=2\pi\frac{\int_0^{+\infty}E(k)/k\,{\rm d}k}{\int_0^{+\infty}E(k){\rm d}k},
\end{equation}
where $k$ is the norm of the wave vector for Snoopy, the norm of the horizontal wave vector for Bala\"itous, and $E(k)$ the kinetic energy spectrum (2D horizontal for Bala\"itous and 3D for Snoopy).
In both cases the turbulent velocity scale $w$ is defined as $\sqrt{2K}$, where $K$ denotes the turbulent kinetic energy per unit mass.

\subsection{Size of the numerical domain}
\label{sec:size}

According to Eq.~\eqref{eq:dtzahn}, the quantity $D_{\rm t}/(\kappa Ri^{-1})$ is constant in Zahn's model.
This is clearly not what is observed in Fig.~\ref{fig:comp}.
Either, one of Zahn's assumptions is not verified, or this is a numerical issue.
In this section, we study the effect of the size of the box on the transport.
In Sect.~\ref{sec:rel}, we focus on the dependence of the transport on the turbulent Reynolds number.

\begin{figure}
    \resizebox{\hsize}{!}{\includegraphics{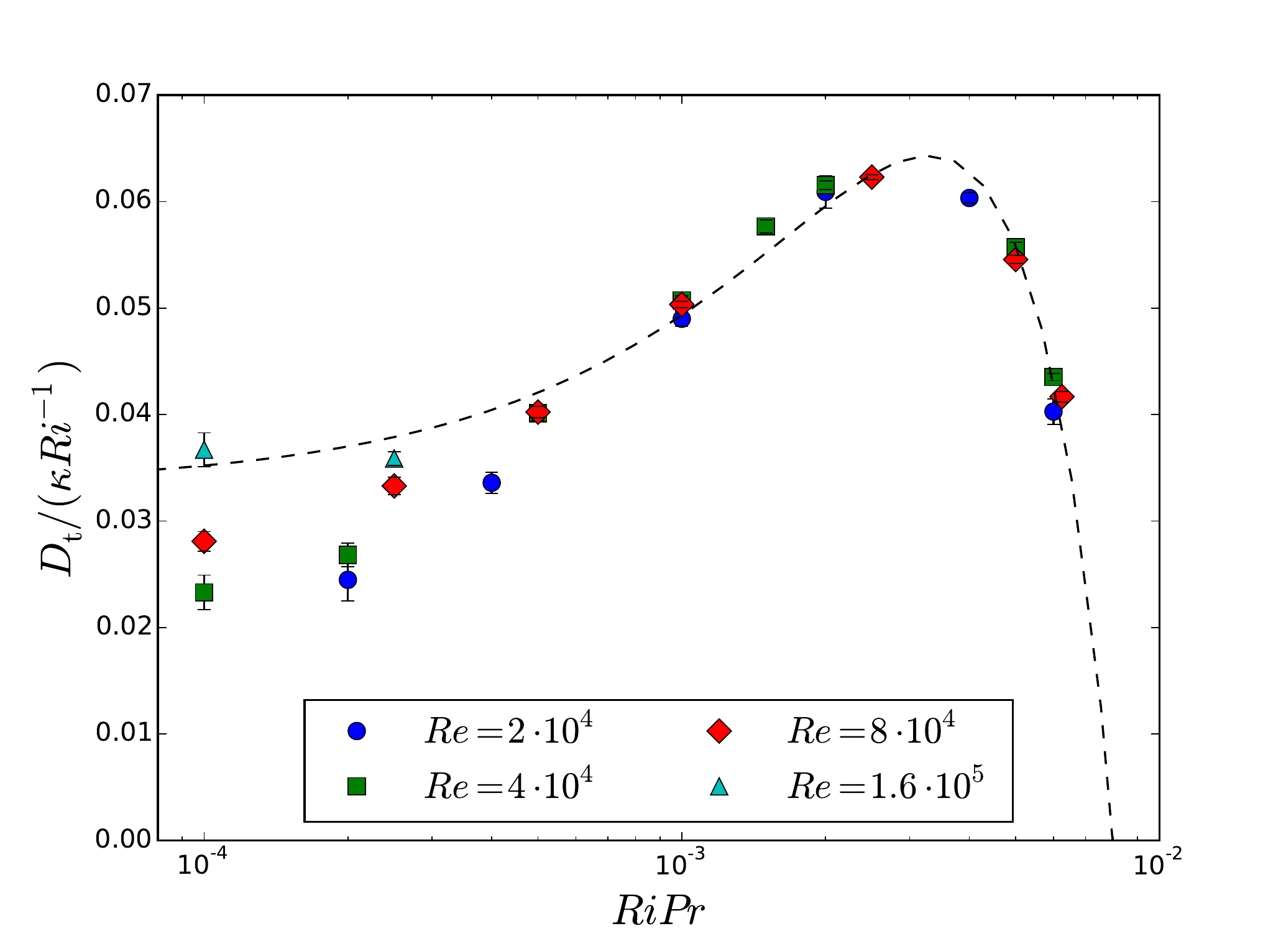}}
    \caption{$D_{\rm t}/(\kappa Ri^{-1})$ as a function of $RiPr$ obtained with our Snoopy simulations. The dotted line corresponds to the empirical formula given in Eq.~\eqref{eq:emp}. There are two overlapping symbols at $RiPr=5\cdot10^{-4}$.}
    \label{fig:ripr}
\end{figure}

The quantity $D_{\rm t}/(\kappa Ri^{-1})$ is plotted in Fig.~\ref{fig:ripr} again as a function of $RiPr$, but this time for all simulations performed with Snoopy.
On the one hand, for values of $RiPr$ down to $5\cdot10^{-4}$, there is no significant difference between the simulations at different Reynolds numbers, so that $D_{\rm t}/(\kappa Ri^{-1})$ can be considered as a function of $RiPr$ only.
On the other hand, for smaller values, a large dispersion in the results appears, larger Reynolds numbers giving larger values of $D_{\rm t}/(\kappa Ri^{-1})$.
As seen previously, $RiPr$ is a proxy for the turbulent Reynolds number, while $Re$ depends on the size of the numerical domain.
Varying $Re$ at constant $RiPr$ is thus equivalent to studying the effect of the numerical domain size.
It means that the results of our simulations depend on the size of the box when $RiPr$ is smaller than $5\cdot10^{-4}$, but not for larger values.

\begin{figure}
    \resizebox{\hsize}{!}{\includegraphics{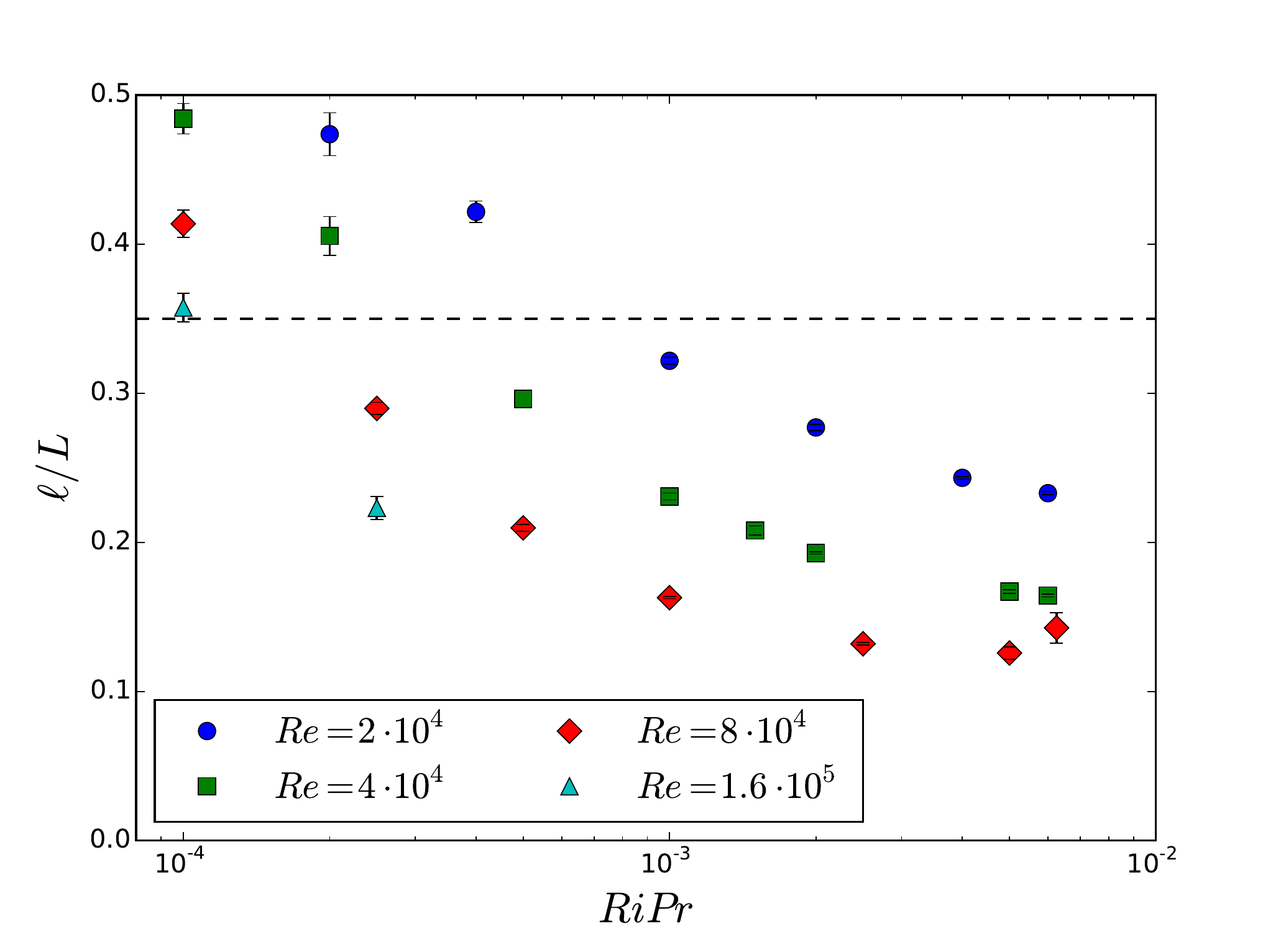}}
    \caption{$\ell/L$ as a function of $RiPr$ obtained with our Snoopy simulations. The dashed line represents the approximate limit $\ell/L=0.35$ between simulation domains that are large enough (below) and too small (above).}
    \label{fig:ell}
\end{figure}

Figure~\ref{fig:ell} shows the ratio $\ell/L$ between the integral scale of turbulence and the size of the numerical domain as a function of $RiPr$.
For simulations with $RiPr\geq5\cdot10^{-4}$, in which the measured transport does not depend on the size of the numerical domain, we observe that the scale ratio $\ell/L$ is always smaller than $0.33$.
Simulations at low $RiPr$ that are clearly affected by the size of the numerical domain show a scale ratio larger than $0.4$.
This suggests that there is a limit in $\ell/L$ at about $0.35$ between simulations which have a sufficiently large box size and those which have a too-small numerical domain.
The latter show a less vigourous turbulence, which can be related to the fact that the size of the numerical domain limits turbulence development.

We now have a reliability criterion which roughly states that there should be at least three large turbulent structures in the simulation domain to be able to give size-independent results.
Assuming that this criterion is valid, two consequences arise.
Firstly, the two simulations at $RiPr=2.5\cdot10^{-4}$ (runs 22 and 30) should not depend on the box size.
Although we can see in Fig.~\ref{fig:ripr} that they have slightly different values of $D_{\rm t}/(\kappa Ri^{-1})$, this difference is significantly smaller than the one between simulations at $RiPr=10^{-4}$ and not much larger than in the regime of larger $RiPr$, for example at $RiPr=10^{-3}$.
Secondly, the simulation with $Re=1.6\cdot10^5$ and $RiPr=10^{-4}$ (run 29) is very close to the limit between size-independent simulations and size-dependent ones.
Therefore, this simulation cannot be completely trusted, but the fact that it gives a value of $D_{\rm t}/(\kappa Ri^{-1})$ similar to the one in the simulation with the same Reynolds number but $RiPr=2.5\cdot10^{-4}$ (run 30) is an argument in favour of its reliability.

To conclude, this criterion should be used in future simulations to ensure that the numerical domain is large enough.
Thanks to this criterion, we can now study the dependence of $D_{\rm t}/(\kappa Ri^{-1})$ on the turbulent Reynolds number, which is detailed in the next section.

\subsection{Turbulent Reynolds number}
\label{sec:rel}

Now that we have determined which simulations are physically relevant, we can focus on the dependence on the turbulent Reynolds number.
If we ignore unreliable simulations (indicated in Table~\ref{tab:res} by a superscript \emph{e}), Fig.~\ref{fig:ripr} shows three main features:
(i) an asymptotic regime at low $RiPr$, which is equivalent to high turbulent Reynolds numbers;
(ii) a maximum around $RiPr=3\cdot10^{-3}$;
and (iii) a very sharp drop around $RiPr=7\cdot10^{-3}$ (simulations above this value showed no turbulence).
Only the asymptotic regime was described by \citet{Zahn}, under the assumption that the turbulent Reynolds number is larger than a critical value of the order of $10^3$.
The two other features are consequences of the transition from quasi-laminar flows dominated by viscosity to fully turbulent flows.
In particular, the lower limit in $Re_\ell$ corresponds to the situation where the largest scale that is unstable with respect to the shear instability is too small to overcome the effect of viscosity.
The fact that this limit can be associated with a critical $RiPr$ has been predicted by \citet{Garaud} using a non-linear stability analysis.

At the moment, we have no physical model for the value of the turbulent diffusion coefficient outside of the asymptotic regime.
However, we can propose the following empirical formula to account for the main features of Fig.~\ref{fig:ripr}:
\begin{equation}
    \label{eq:emp}
    \frac{D_{\rm t}}{\kappa Ri^{-1}} = \alpha+\beta RiPr - \gamma(RiPr)^2,
\end{equation}
where $\alpha=3.34\cdot10^{-2}$, $\beta=18.8$, and $\gamma=2.86\cdot10^3$.
In the asymptotic regime (i.e. for small values of $RiPr$), we thus have $D_{\rm t}/(\kappa Ri^{-1})\simeq\alpha$.
The value we find here is almost twice as low as the value $5.58\cdot10^{-2}$ found in PL13 because the asymptotic regime was not reached in that paper.
The physical meaning of $\beta/\gamma=6.57\cdot10^{-3}$ is the following: for $RiPr>\beta/\gamma$, $D_{\rm t}/(\kappa Ri^{-1})$ drops below the asymptotic value.
Since the curve is quite steep, $\beta/\gamma$ would effectively give the order of magnitude of the critical $RiPr$ above which there is no turbulence.

In the asymptotic regime, our new prescription predicts about 60\% less transport than the model by \citet{Zahn}, and about 15 times less than the model by \citet{Maeder1995}.
Outside of the asymptotic regime, the discrepancy between our prescription and those by Zahn and Maeder can be reduced at best by a factor two.
Even then, our prescription predicts still significantly less transport than the model by \citet{Maeder1995}.

\begin{figure}
    \resizebox{\hsize}{!}{\includegraphics{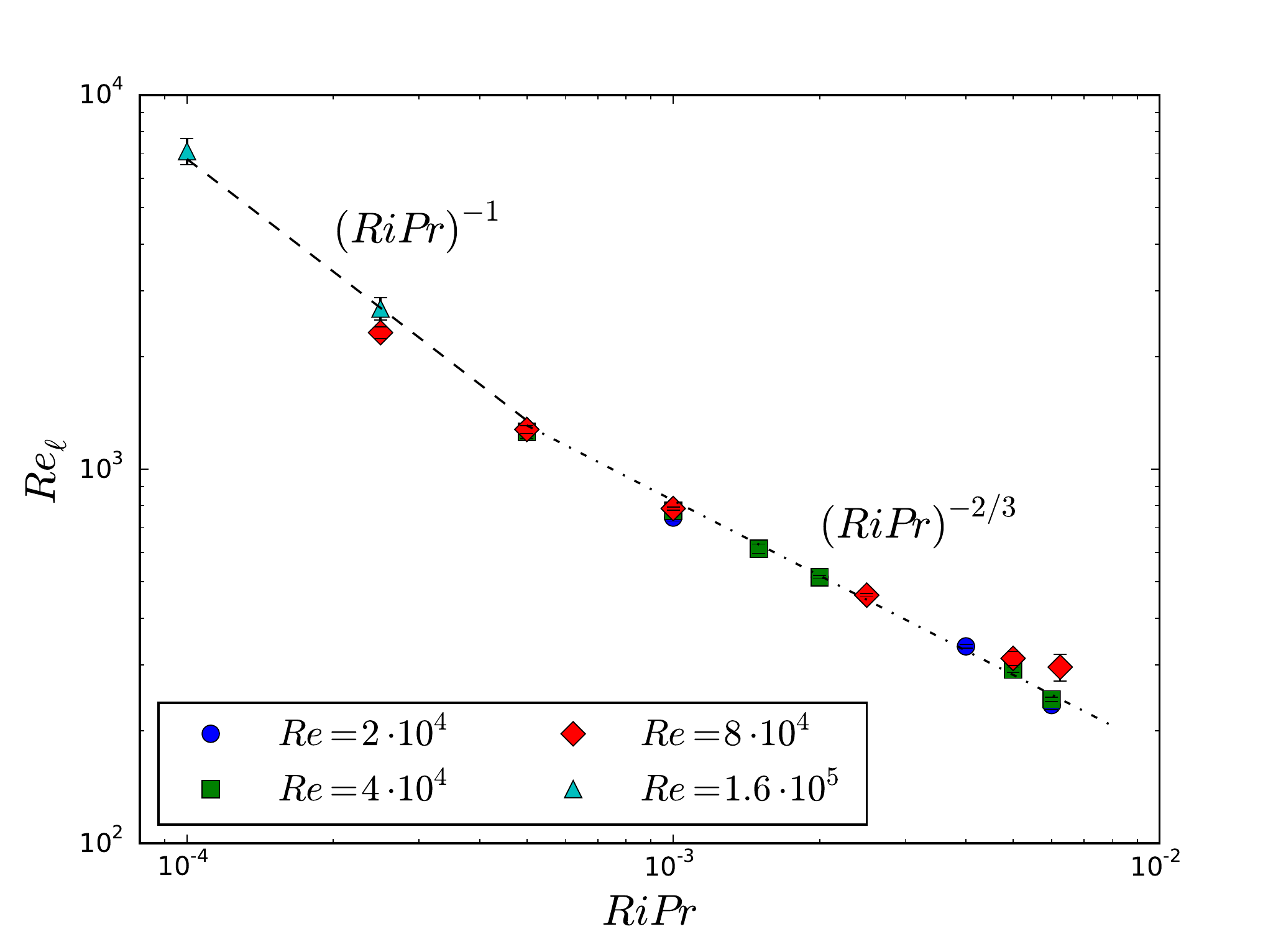}}
    \caption{$Re_\ell$ as a function of $RiPr$ for all valid Snoopy simulations. The dashed line and the dash-dotted line represent the power laws $Re_\ell \propto (RiPr)^{-1}$ and $Re_\ell \propto (RiPr)^{-2/3}$, respectively.}
    \label{fig:rel}
\end{figure}

In the following figures, we kept only the relevant simulations.
The relation between $Re_\ell$ and $RiPr$ (see Eq.~\eqref{eq:rel_ripr}) is tested in Fig.~\ref{fig:rel}.
We observe that Eq.~\eqref{eq:rel_ripr} is verified only when $RiPr<5\cdot10^{-4}$, that is close to the asymptotic regime (dashed line in Fig.~\ref{fig:rel}).
This coincides quite well with the condition $Re_\ell>10^3$ given by \citet{Zahn}. 
Even at larger values of $RiPr$, the relation between $Re_\ell$ and $RiPr$ is always a one-to-one relationship.

\begin{figure}
    \resizebox{\hsize}{!}{\includegraphics{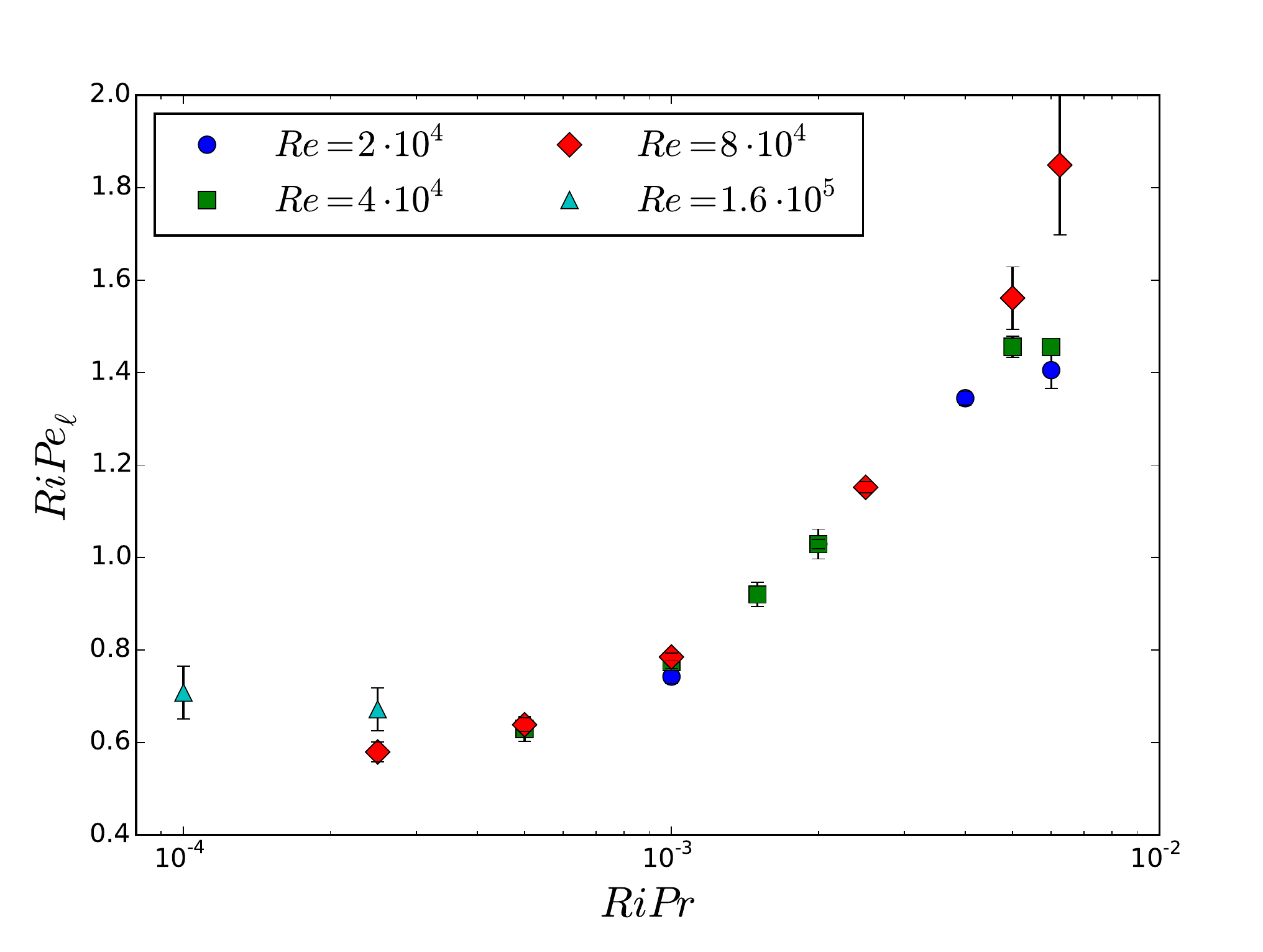}}
    \caption{$RiPe_\ell$ as a function of $RiPr$ for all valid Snoopy simulations.}
    \label{fig:ripel}
\end{figure}

The fact that Eq.~\eqref{eq:rel_ripr} is verified for low values of $RiPr$ implies that $RiPe_\ell$ is almost constant in this regime.
This is illustrated in Fig.~\ref{fig:ripel}.
We can see that the value of $RiPe_\ell$ is close to $0.65$ in the considered regime.
This verifies directly the assumption of Zahn's model that in the asymptotic regime flows are characterised by a constant value of the turbulent Richardson-P\'eclet number.
The approximate value found here is about 50\% larger than $0.426$, the value found in PL13.
As mentioned earlier, this may result from the fact that the integral scale $\ell$ was somewhat differently defined in previous studies (PL13 and Paper I) compared to the present one, while the definition of $w$ remains the same.
We observe that the values of $\ell$ are larger in the present paper, and so are $Re_\ell$ and $RiPe_\ell$.
This may also be partly due to the fact that the SPNA simulation in PL13 was not in the asymptotic regime. 

\begin{figure}
    \resizebox{\hsize}{!}{\includegraphics{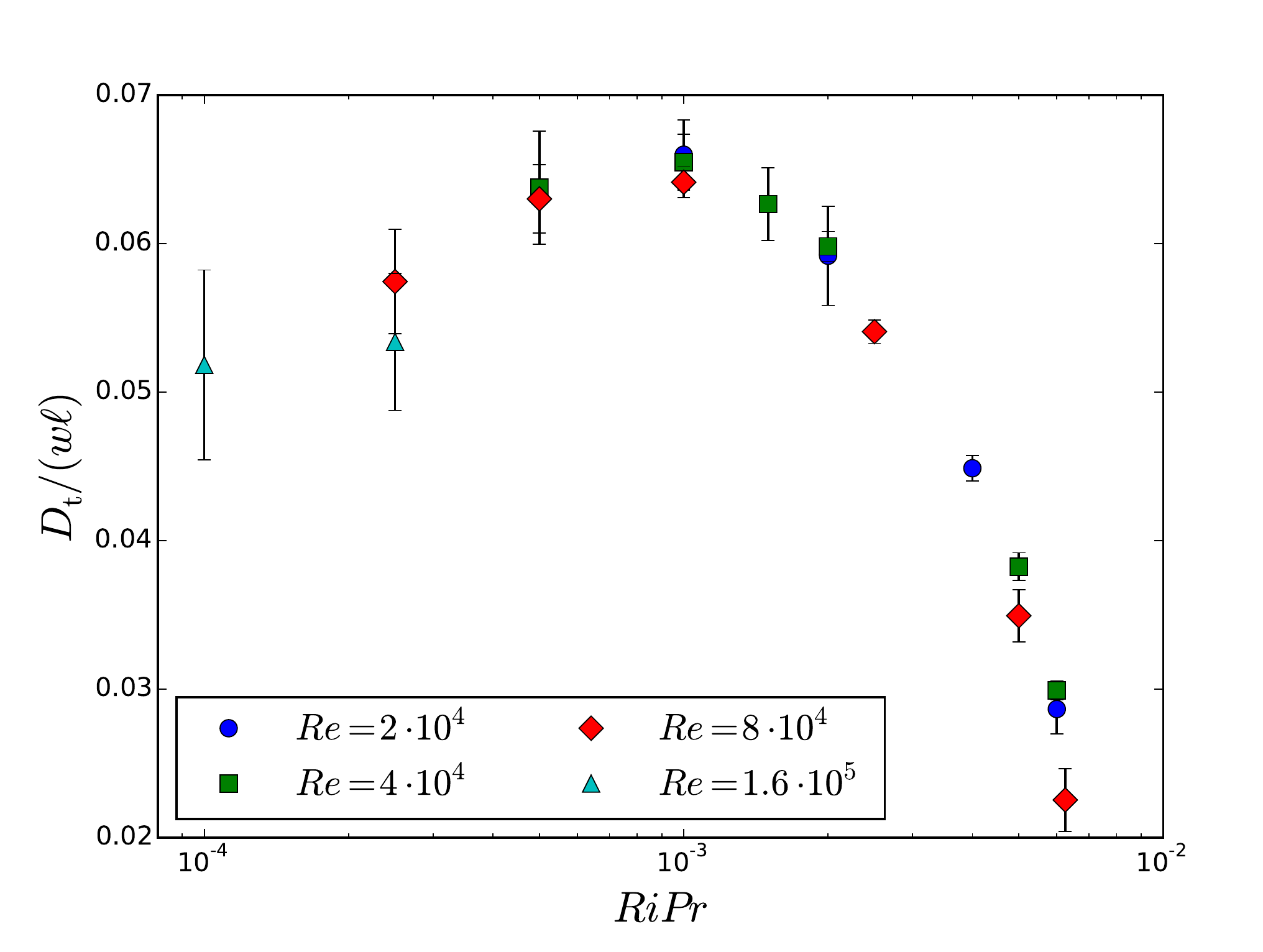}}
    \caption{$D_{\rm t}/(w\ell)$ as a function of $RiPr$ for all valid Snoopy simulations.}
    \label{fig:dtsul}
\end{figure}

Another of Zahn's assumptions can be tested with our simulations: the fact that the turbulent diffusion coefficient is proportional to $w\ell$.
We observe in Fig.~\ref{fig:dtsul} that the ratio $D_{\rm t}/(w\ell)$ is close to $5.5\cdot10^{-2}$ when $RiPr\leq2.5\cdot10^{-4}$, then increases up to $6.5\cdot10^{-2}$ and finally drops to much smaller values, the value in the asymptotic regime being significantly smaller than $0.131$, the one found in PL13.
Again, this is probably due to the different definitions of $\ell$.
We can also see that the assumption that $D_{\rm t}/(w\ell)$ is constant is more restrictive than the previous one, since it seems to be true only for smaller values of $RiPr$. 

\subsection{Relation between turbulent diffusion and viscosity coefficients}
\label{sec:nul}

\begin{figure}
    \resizebox{\hsize}{!}{\includegraphics{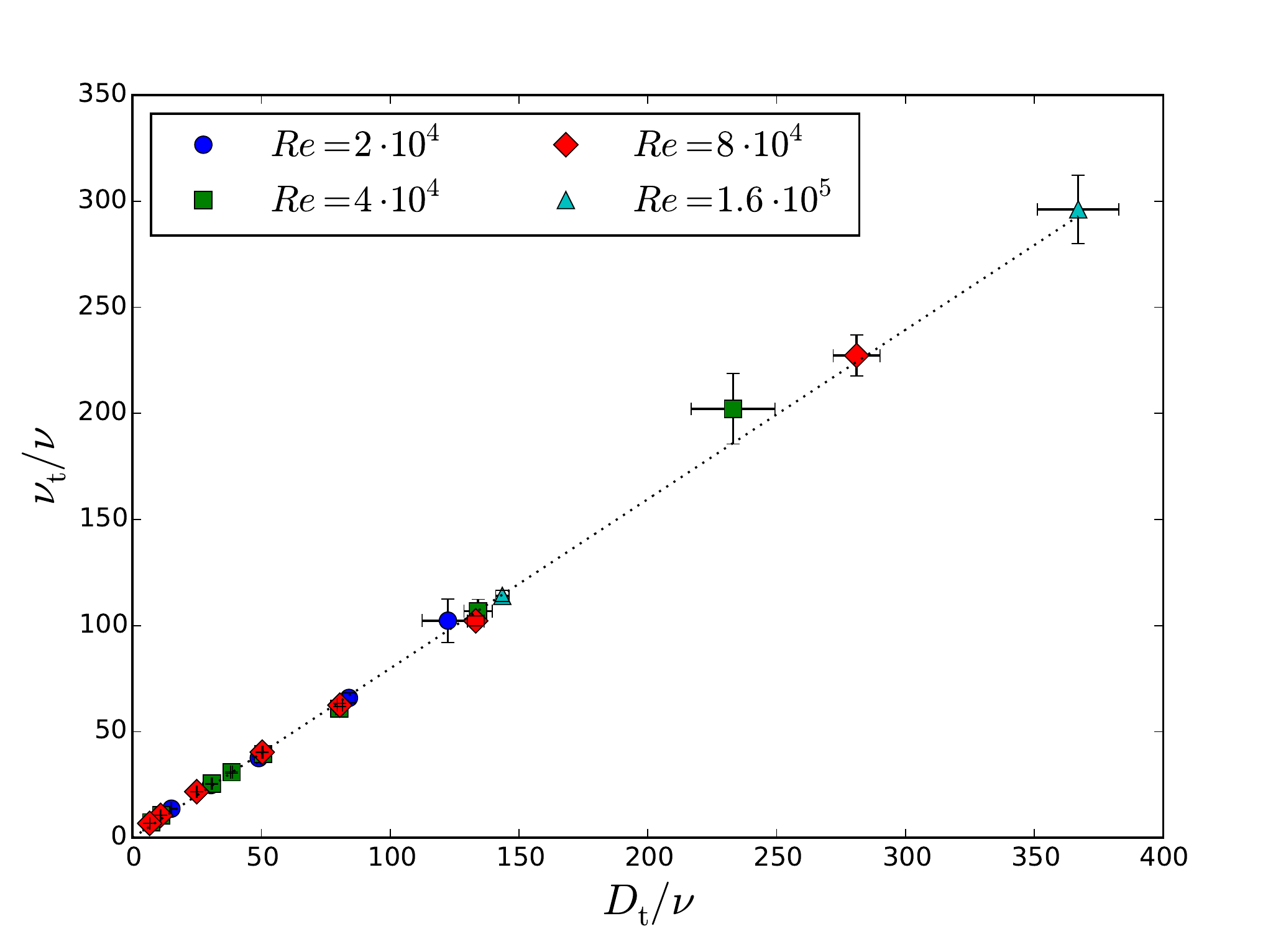}}
    \caption{Scaling between $\nu_{\rm t}$ and $D_{\rm t}$ for all Snoopy simulations. The dotted line represents the proportionality relation $\nu_{\rm t}=0.80 D_{\rm t}$.}
    \label{fig:nut}
\end{figure}

Many stellar evolution codes consider that the diffusion coefficients for chemical elements and for angular momentum are the same, even though \citet{Zahn} mentioned that they could differ by a small factor.
In the weakly turbulent regime ($RiPr\gtrsim 3\cdot10^{-3}$), our simulations show that both coefficients are small, and have nearly equal values ($D_{\rm t}\sim\nu_{\rm t}\lesssim10\nu$).
In the fully turbulent regime, which is relevant for stellar interiors, we find that they follow the relation
\begin{equation}
    \nu_{\rm t} \simeq 0.80 D_{\rm t},
\end{equation}
as illustrated in Fig.~\ref{fig:nut}.

\section{Discussion}
\label{sec:discuss}

We have shown that the two codes we have tested give similar results in the regime of small P\'eclet numbers.
This provides a validation of the shearing-box formalism for the study of the shear instability in this regime.
Our work suggests further that we do not always have a clear scale-separation between the turbulent dissipation scale, the largest shear-unstable scale, and the size of the computational domain.
As a consequence, one must be cautious in the interpretation of the simulation results.
In particular, we showed that it is crucial to have a numerical domain at least three times as large as the largest shear-unstable scale to ensure that the results are relevant.
Our results also suggest that the asymptotic regime described by \citet{Zahn} is reached when the turbulent Reynolds number is larger than $10^3$.
This is in agreement with the fourth of Zahn's assumptions mentioned in Sect.~\ref{sec:zahn}.
Such high values of the turbulent Reynolds number were not reached in our previous studies (PL13 and Paper I), which were consequently not in the asymptotic regime.

The observed dependence of the turbulent diffusion on the turbulent Reynolds number (or on $RiPr$, which is easier to calculate in stellar evolution codes) is not trivial.
The empirical prescription we propose in this paper can be implemented in stellar evolution codes and compared with other prescriptions.
In addition, we have shown that it is necessary to distinguish between the turbulent viscosity coefficient used for the transport of angular momentum and the turbulent diffusion coefficient used for the mixing of chemical elements.
In practice, the proportionality constant between them is close to one, which means its impact on stellar evolution calculations is expected to be minor.
The fact that our new prescriptions predicts even less transport than in our previous studies does not reduce the disagreement between observations and stellar evolution models.

The validation of Snoopy in this context is an interesting result, since it is able to simulate configurations in which the velocity and entropy gradients are not aligned.
Such configurations would allow us to study the horizontal transport generated by some horizontal shear and the effect of this horizontal transport on the vertical transport \citep[see model by][]{TalonZahn}.
The horizontal transport, believed to weaken the effect of the stable stratification and thus enhance the vertical transport, is indeed poorly understood.
Several prescriptions exist \citep[see for example][]{Zahn,Maeder,Mathis}, but all are based on phenomenological arguments and have never been tested.
Constraints from numerical simulations are therefore needed.
One must keep in mind that the similarities in the behaviour of the two codes are not necessarily valid in the regime of large P\'eclet numbers.
In this regime, the boundary conditions and the forcing scheme probably play an important role, since in the classical Richardson stability criterion, $Ri > Ri_{\rm crit}$ , no preferential physical scale is involved.
To give reliable prescriptions of turbulent transport in this regime, which may occur in evolved stars \citep{Hirschi}, the impact of the differences between the two configurations must be investigated in detail.

To explore further the asymptotic regime, we must go to higher Reynolds numbers, which means better grid resolutions, but also larger box sizes, thus increasing drastically the computational cost.
As mentioned in the introduction, the expected Reynolds number in stellar radiative zones is between $10^2$ and $10^5$.
Therefore, our simulations already overlap with the physically relevant regime for stellar interiors.
Although it requires significantly more computational effort to reach Reynolds numbers of the order of $10^5$, such computations should be possible in the near future thanks to the continuous increase in available computational power.
We note that this favourable situation contrasts with other problems in stellar hydrodynamics, for example global simulations of the Solar convective region, which are hampered by the very large Reynolds number (of the order of $10^9$) that characterises the problem, making fully resolved simulations unfeasible in the foreseeable future.
For the shear problem, a cheaper alternative would be to perform large-eddy simulations (LES) in which small scales are modelled using prescriptions from DNS.
But before doing so, one would first have to test this approach by comparing LES and DNS with the same control parameters.

For simplicity, the dynamical effect of chemical stratification was not taken into account in the present study, but the respective results found in Paper I are very likely to also depend on the turbulent Reynolds number.
This will be the subject of a forthcoming publication.
Finally, another potentially relevant ingredient is the magnetic field.
In the presence of differential rotation, the magneto-rotational instability (MRI) may drive MHD turbulence and therefore provide further transport \citep{Balbus, Spruit, Menou, Wheeler, Jouve}.
The MRI in a stably stratified layer has so far mostly been described analytically, through linear analysis and phenomenological arguments.
These results need to be checked with numerical simulations in a similar way to that in the present paper.
Analogous simulations have been performed in the context of proto-neutron stars \citep{Guilet} and should be extended to the regime of small P\'eclet numbers that is relevant for stellar radiative zones.

\begin{acknowledgements}
VP thanks Geoffroy Lesur for his interesting ideas and his help in the use of the Snoopy code.
VP and MV thank Fran\c cois Ligni\`eres for insightful discussions.
VP acknowledges support from the European Research Council through ERC grant SPIRE 647383.
JG acknowledges support from the Max-Planck-Princeton Center for Plasma Physics.
This work is supported by the European Research Council through grant ERC-AdG No.~341157-COCO2CASA.
The authors thank the referee Georges Meynet for his useful comments that contributed to the quality of the paper.
\end{acknowledgements}

\bibliographystyle{aa}
\bibliography{refs}

\end{document}